\documentstyle[prl,aps]{revtex} 
\begin{document}
\draft
\title{ Exact results of a hard-core interacting system with
a single impurity}
\author{You-Quan Li$^{1,2}$ and Zhong-Shui Ma$^{1,3}$ }
\address{
${}^{1}$CCAST(World Laboratory), P.O. Box 8730 Beijing 100080, China\\
\footnote{mailing address}${}^{2}$Zhejiang Institute of Modern Physics,
Zhejiang University, Hangzhou 310027, China \\
${}^{3}$Advanced Research Center, Zhongshan University,
Guangzhou 510275, China
}
\date{\today}
\maketitle
\begin{abstract}
The quantum-mechanical problem of a many-particle system with
a single impurity in one-dimension, interacting by a delta-function,
is solved. The wave-function for a bosonic system and the related secular
equation for the spectrum are obtained.
The energy per length is calculated in the thermodynamic limit.
\end{abstract}

\narrowtext
\section*{}

The quantum-mechanical problem of particles in one-dimension with
delta-function interaction was studied in [1-5]. In \cite{Yang},
the strategy now referred to as `Bethe ansatz' \cite{Bethe} was
successfully applied to this problem. In order
to obtain the secular equation for the spectrum, periodic boundary condition
is imposed so that the particles were thought of as situated on a circle.
Subsequently, some other kind of boundary conditions were taken into account,
such as the case of particles being enclosed in an infinitely deep potential 
well \cite{Gaud2}, the case of \cite{Gaud2} with a modification by addition
of an infinitely deep and very narrow well at one end of the interval
\cite{Woyn} or, recently, the case of particles being enclosed in a well
of finite depth \cite{LYQ}. As a mathematical consideration, \cite{Suthld2}
had generalized the above problem to the extent within  finite Coxeter group.

In the series of a line of studies on exact solutions of the above mentioned
quantum-mechanical problem, as far as we are aware, attention has largely
been concentrated on the replacement of boundary conditions which are
used to determine the secular equation. In the present letter we consider
a system of $N$-particles and a single impurity. We find the Bethe ansatz
wave-function of the system and the secular
equation for the spectrum. As a result, the wave-function is a sum of
non-diffractive waves over  the permutation group $S_{N}$ and defined 
piecewisely on separate regions. These regions  are specified by the
permutation group $S_{N+1}$ instead of $S_{N}$. The energy per length
of the system is calculated in the thermodynamic limit. The problem
involved in the present letter differs from that of the  Kondo Hamiltonian
which was ever diagonalized in \cite{Andr,Wgman}.

The Hamiltonian of a system of $N$-particles and a single impurity,
interacting by a delta-function, reads
\begin{equation}
H = - \frac{\hbar^{2}}{2m}\sum^N_{i=1}\frac{\partial^2 }{\partial x_{i}^2 }
- \frac{\hbar^{2}}{2m'}\frac{\partial^2 }{\partial x'^2}
+ u\sum^N_{j>i=1}\delta(x_{i} - x_{j})
+ v\sum^N_{i=1}\delta(x_{i} - x')
\label{eq:a}\end{equation}
where $x'$ stands for the coordinate of the impurity and $x_{i}$ for that of
the $i$th particle. For simplicity, we start discussion with the convention
$\mu := m'/m, \hbar=1, m=1$. The Hamiltonian (\ref{eq:a}) is invariant
under the action of whole coordinates translation, so the total momentum
of the system is conserved.

The continuity of wave-function  and their derivatives is determined by
the Schr\"{o}dinger equation on the basis of the properties of the
potential in the Hamiltonian.
In order to conveniently obtain the boundary condition arising from the
delta-function terms in (\ref{eq:a}), we make a scalar transformation
$ x' \rightarrow x_{0} =\sqrt{\mu}x'$.
With this transformation, we can combine the first and second terms of
(\ref{eq:a}) so that they become a standard Laplace operator in an $(N+1)$
dimensional euclidean space ${\sf \, l \! R}^{N+1}$ with cartesian
coordinates
$(x_{0}, x_{1}, x_{2},\cdots, x_{N})$.
Thus, the Schr\"{o}dinger equation takes the following form
\begin{equation}
\left[ -\frac{1}{2}\sum_{l=0}^{N}\frac{\partial^2 }{\partial x_{l}^2 }
+ u\sum^N_{j>i=1}\delta(x_{i} - x_{j})
+ v\sum^N_{i=1}\delta(x_{i} - \frac{1}{\sqrt{\mu}}x_{0} ) \right]
\psi (x_{0}, x_{1}, x_{2}, \cdots, x_{N})
= E \psi (x_{0}, x_{1}, x_{2}, \cdots, x_{N})
\label{eq:b}
\end{equation}
where the summation of the `kinetic energy' starts from $l=0$ instead of
$l=1$.

After removing the delta-function terms from the left hand side to the right
hand side of (\ref{eq:b}), one takes integration of (\ref{eq:b})
over  a `Gauss box'
that is cut into halves by  either the hyperplane
$\{x_{i} -x_{j} = 0 | \, i=1, 2, \cdots, N \}$ or
$\{x_{i} - x_{0}/\sqrt{\mu} =0 | \, i=1, 2,\cdots, N \}$.
With the help of Gauss integral theorem, we obtain the discontinuity
relations 
for the derivatives of  wave-function along the normal of a hyperplane
${\sf \, l \! P}_{\alpha}=\{x |\, (\alpha | x)=0 \} $, where $\alpha$
stands for
the normal vector and $(\alpha | x )$ stands for a scalar product of two 
vectors $\alpha$ and 
$x =(x_{0}, x_{1}, x_{2},\cdots, x_{N})$ i.e.
\begin{equation}
\lim_{\epsilon \rightarrow 0^{+}}\{ \alpha\cdot\nabla[
\psi (x_{(\alpha)} + \epsilon \alpha) -
\psi (x_{(\alpha)} - \epsilon \alpha ) ] \}
= 2c \psi(x_{(\alpha)})
\label{eq:c}\end{equation}
where $c=u$ for $\alpha = e_{i} -e_{j}$
and $c=v$ for $\alpha = e_{i}-\displaystyle{\frac{1}{\sqrt{\mu}}}e_{0}$
$( i=1, 2,\cdots, N)$;
$\nabla:= \displaystyle
{\sum_{l=0}^{N}} e_{l}\frac{\partial }{\partial x_{l}}$;
$x_{(\alpha)}\in {\sf \, l \! P}_{\alpha}$ and $\{ e_i \}$ is the standard
orthogonal basis.

It is convenient to discuss  the present problem in the frame of reference
of the impurity. This requires a coordinate transformation
$(x_{0}, x_{1}, x_{2},\cdots, x_{N}) \mapsto $
$(y_{0}, y_{1}, y_{2},\cdots, y_{N})$
given by
$y_{0}:= x' =\displaystyle{\frac{1}{\sqrt{\mu}}}x_{0},$ and
$y_{i}:=x_{i} - x' = x_{i}-\displaystyle{\frac{1}{\sqrt{\mu}}}x_{0}$.
It results that
$\displaystyle{
\frac{\partial }{\partial x_{0}} = \frac{1}{\sqrt{\mu}}
(\frac{\partial }{\partial y_{0}}
- \sum^N_{i=1}\frac{\partial }{\partial y_{i}})
}$ and
$\displaystyle{
\frac{\partial }{\partial x_{i}} = \frac{\partial }{\partial y_{i}}
}$. Obviously,
$\displaystyle{
\frac{\partial }{\partial y_{0}} = \sum^N_{i=1}
\frac{\partial }{\partial x_{i}} + \frac{\partial }{\partial x'}
}$ is just the operator of total momentum (apart from a factor $-i$).
In terms of these coordinates
$ (y_0, y_1, y_2,\cdots, y_N )$,
the Hamiltonian (\ref{eq:a})
and the discontinuity relation (\ref{eq:c}) become
\begin{equation}
H= -\frac{1}{2}\sum^N_{i=1}\frac{\partial^2 }{\partial y_{i}^2 }
-\frac{1}{2\mu}\left(\frac{\partial }{\partial y_{0}}
-\sum^N_{i=1}\frac{\partial }{\partial y_{i}}\right)^{2}
+u\sum^N_{j>i=1}\delta(y_{i}-y_{j}) + v\sum^N_{i=1}\delta(y_{i}),
\label{eq:d}\end{equation}
\begin{equation}
\lim_{\epsilon\rightarrow 0^{+}}\left(\frac{\partial }{\partial y_{i}}
-\frac{\partial }{\partial y_{j}}\right)
\left[ \psi(y_{(\alpha)} +\epsilon\alpha)-\psi(y_{(\alpha)} -\epsilon\alpha)
\right] = 2u \psi (y_{(\alpha)} ),
\label{eq:e}\end{equation}
and
\begin{equation}
\lim_{\epsilon\rightarrow 0^{+}}
\left[\frac{\partial }{\partial y_{i}}-\frac{1}{\mu }
(\frac{\partial }{\partial y_{0}}-
\sum^N_{i=1}\frac{\partial }{\partial y_{i}})\right]
\left[ \psi(y_{(\alpha')} +\epsilon\alpha')-\psi(y_{(\alpha')}
-\epsilon\alpha'
)\right] = 2v \psi (y_{(\alpha')} ),
\label{eq:f}\end{equation}
where
$\alpha$
stands for the normal vector of the hyperplane
$\{y |\, y_{i}-y_{j} =0 \}$ and $\alpha' $
for that of the hyperplane
$\{y |\, y_{i} = 0 \}$.
Clearly, (\ref{eq:d}) is invariant under the translation of
$y_{0} \rightarrow y_{0} + \epsilon $ 
or under any permutation of the coordinates 
$y_{i}$ for $i=1, 2,\cdots, N$.
Because of the translational invariance of $y_{0}$, we may set
$
\psi (y_{0}, y_{1},\cdots, y_{N})
= e^{iK y_{0}}\varphi(y_{1},\cdots, y_{N})
$,
where $K$ is a constant, i.e. the total momentum is conserved. Then the
Schr\"{o}dinger equation is reduced to a differential equation for
$\varphi (y_{1}, y_{2},\cdots, y_{N})$. On the domain
${\sf \, l \! R}^{N}\setminus \{{\sf \, l \! P}_{\beta} \}$ it becomes
\begin{equation}
\left[ - \frac{1}{2}\sum^N_{i=1}\frac{\partial^2 }{\partial y_{i}^2 }
-\frac{1}{2\mu }( \sum^N_{i=1}\frac{\partial }{\partial y_{i}}
-iK \,)^{2} \right] \varphi (y_{1}, y_{2},\cdots, y_{N})
= E \varphi (y_{1}, y_{2},\cdots, y_{N} ),
\label{eq:g}\end{equation}
where
$\{ {\sf \, l \! P}_{\beta} \}$
stands for the set of hyperplanes
$\{y |\, y_{i} -y_{j} = 0 \}$ and
$\{y |\, y_{i}=0 \}$ for $i, j = 1, 2,\cdots,  N$.
Obviously these hyperplanes partition
${\sf \, l \! R}^{N}$
into finitely many regions and a wave-function must be piecewisely continuous
function defined respectively on separate regions. The regions into which
the hyperplanes
$\{ \{y| \, y_{i} -y_{j} =0 \} | i, j=1, 2,\cdots, N \} \}$
partition ${\sf \, l \! R}^{N}$  can be specified by elements of
a permutation group
$S_{N}$.
In other words, different regions have  different orders for the coordinates
$(y_{1}, y_{2},\cdots, y_{N} )$ 
if their components are arranged from large value to small ones, e.g.
$y_{1} > y_{2} > y_{3} >\cdots > y_{N}$, \,
$y_{2} > y_{1} > y_{3} >\cdots > y_{N}$
etc. Moreover, each of such regions is partitioned into
$N + 1$ separate regions by the other type of hyperplanes
$\{ \{y | \,y_{i} =0 \} \}$, For example, the region
$y_{1} > y_{2} >\cdots > y_{N}$,
is partitioned into the regions
$0 > y_{1} > y_{2} >\cdots > y_{N}$,
$y_{1} > 0 > y_{2} >\cdots > y_{N}, \cdots $ and
$y_{1} > y_{2} >\cdots > y_{N} > 0$.
These regions can be specified by a permutation group $S_{N+1}$.
It is easy to find a
special solution for (\ref{eq:g}), i.e. a plane wave solution
$\varphi_{\{ k \} }(y) \sim e^{ i( k | y )}$, here
$ ( k | y ) = k_{1}y_{1} + k_{2}y_{2} + \cdots + k_{N}y_{N} $.
The corresponding energy is 
$E =\displaystyle{ \frac{1}{2}\sum^N_{i=1}k_{i}^{2}
+ \frac{1}{2\mu }( \sum^N_{i=1} k_{i} -K )^{2} }$
or more neatly,
\begin{equation}
E = \frac{1}{2}\sum^N_{i=1}k_{i}^{2} + \frac{1}{2\mu }\lambda^{2}
\label{eq:h}\end{equation}
where a parameter $\lambda = K -\sum^N_{i=1}k_{i}$ is introduced.

Because of the  permutational invariance  of the Hamiltonian (\ref{eq:d}),
we look for solutions of the Bethe ansatz form:
\begin{equation}
\varphi_{\tau }(y) = \sum_{\sigma \in S_{N} } A(\sigma, \tau ) 
e^{i(\sigma k | y ) },
\label{eq:i}\end{equation}
where $\sigma k $ stands for the image of a given 
$k:= (k_{1}, k_{2}, \cdots, k_{N})$
by a permutation 
$\sigma \in S_{N}$ and the coefficients
$A(\sigma, \tau )$ are functionals on
$S_{N} \otimes S_{N+1}$. It is worthwhile to mention that the summation in
(\ref{eq:i}) is taken over the permutation group $S_{N}$ whereas
the various regions
on which the wave-function is defined are specified by elements of
$S_{N+1}$. This is different from the Bethe-Yang ansatz \cite{Yang}.
As far as the Bethe ansatz solution (\ref{eq:i}) is concerned, the
boundary condition (\ref{eq:f}) becomes
\begin{equation}
(\frac{\partial }{\partial y_{i}} - i\frac{\lambda }{\mu })
\left[ \varphi (y)|_{ y_{i} \rightarrow 0^{+}}
- \varphi (y)|_{y_{i}\rightarrow 0^{-}} \right]
= 2v\varphi (y) |_{y_{i} \rightarrow 0 }.
\label{eq:j}\end{equation}

For a bosonic system, the wave-function is supposed to be symmetric under
any permutation of the coordinates. Because any element of permutation group
$S_{N} $ can be expressed as a product of the neighboring interchanges
$\sigma_{i}$:
$(\cdots, y_{i}, y_{i+1},\cdots )\mapsto $
$(\cdots, y_{i+1}, y_{i},\cdots ) $,
$ i=1, 2,\cdots,N-1$, it follows that
$(\sigma_{i} \varphi )(y)=\varphi (y)$.
Since  $\varphi $ is a scalar function,
$(\sigma_{i} \varphi )$ is well defined by 
$\varphi (\sigma^{-1}_{i} y )$. Thus both sides of the equation
can be written out by using (\ref{eq:i}). In terms of the evident identity
$(\sigma k | \sigma_{i}^{-1}y ) = ( \sigma_{i} \sigma k | y )$
and the rearrangement theorem of group theory, we obtain the following
consequence:
\begin{equation}
A(\sigma, \sigma_{i}\tau ) = A(\sigma_{i}\sigma, \tau ), \, \,{\rm for} \,
i= 1, 2,\cdots, N-1.
\label{eq:k}\end{equation}
Substituting (\ref{eq:i}) into the boundary condition (\ref{eq:j}) and
using the continuity condition for the wave-function on the hyperplanes,
we obtain the following relation (when $y_{i}$ is nearest to zero
among all $\{y_{j}, j=1,2, \cdots, N \}$)
\begin{equation}
A(\sigma, \sigma_{N}\tau ) 
= - \frac{v-i[(\sigma k)_{i}-\lambda /\mu ]}
         {v+i[(\sigma k)_{i}-\lambda /\mu ]}
A(\sigma, \tau ), 
\label{eq:l}\end{equation}
where $\sigma_{N} \in S_{N+1}$ but $\not\in S_{N}$.
The relations (\ref{eq:k})
and (\ref{eq:l}) relate $A$-coefficients  between different
regions, so we direct our attention to one of the regions, e.g.
$y_{1} > y_{2} >\cdots y_{N} >0$.

Because the  region with
$\cdots y_{i}< y_{i+1} \cdots $ is next to the region with
$\cdots y_{i+1}< y_{i} \cdots $ for $y_{i}y_{i+1} > 0$ 
($\cdots y_{i}<0< y_{i+1} \cdots $ is not next to the region
$\cdots y_{i+1}< 0 < y_{i} \cdots $), the boundary
condition (\ref{eq:e}) and the continuity condition for the wave-function
on the related hyperplanes give a relation between
$\varphi_{\tau }(y)$ and $\varphi_{\sigma_{i}\tau }(y)$
($i=1, 2,\cdots, N-1$).
After writing out the relation in detail we obtain the following:
\begin{equation}
i[ (\sigma k )_{i} - (\sigma k )_{i+1} ] [A(\sigma, \tau )-A(\sigma_{i}\sigma,
\tau) -A(\sigma, \sigma_{i}\tau ) + A(\sigma_{i}\sigma, \sigma_{i}\tau ) ]
= 2u [ A(\sigma, \tau ) + A(\sigma_{i}\sigma, \tau ) ].
\label{eq:m}\end{equation}
With the help of (\ref{eq:k}) this relation gives
\begin{eqnarray}
A(\sigma_{i}\sigma, \tau ) = - Y_{i}(\sigma k ) A(\sigma, \tau ),\nonumber \\
Y_{i}(\sigma k) = \frac{u -i [(\sigma k )_{i} - (\sigma k )_{i+1} ]}
{u + i [(\sigma k )_{i} - (\sigma k )_{i+1} ]}.
\label{eq:n}\end{eqnarray}
As a result, all the $A$-coefficients are determined up to an overall scalar
factor by using (\ref{eq:n}) repeatedly. The consistency of successive use
of (\ref{eq:n}) is guaranteed because the $Y$'s in (\ref{eq:n})
satisfy Yang-Baxter equation.

In the above we have obtained the Bethe ansatz solution of Schr\"{o}dinger
equation on the basis of boundary condition arising from delta-functions in
Hamiltonian. Now we turn to determine the secular equation of the spectrum
$\{ k\}$. Suppose the system is situated on a circle of length $L$, the 
wave-function must obey the periodic boundary condition
$\varphi (y_{1},\cdots, y_{i}-L, \cdots, y_{N})=$
$\varphi (y_{1},\cdots, y_{i},\cdots, y_{N})$. If
$y=(y_{1},\cdots, y_{i},\cdots, y_{N})$
is a point in the region specified by
$\tau \in S_{N+1}$, as a result of periodicity, 
$y' =(y_{1},\cdots, y_{i} - L, \cdots, y_{N} )$ must be a point on another 
region specified by $\gamma\tau \in S_{N+1}$, where
$\gamma = \sigma_{N}\Delta $ and
$\Delta := \sigma_{N-1}\cdots \sigma_{2}\sigma_{1} \in S_{N} $.
Then the requirement of the periodic boundary condition is explicitly
written as
$\varphi_{\tau }(y) = \varphi_{\gamma\tau }(y')$. 
After writing this relation in terms of the Bethe ansatz
solution (\ref{eq:i}), we find that the periodic boundary condition is
guaranteed as long as the  relation
$A(\sigma, \gamma\tau ) e^{-i(\sigma k )_{1} L } = A(\sigma, \tau )$
is imposed. This implies the following equation after using (\ref{eq:k}),
(\ref{eq:l}) and
(\ref{eq:n}) repeatedly,
\begin{equation}
e^{i(\sigma k)_{1}L} = (-1)^{N} \frac
{v - i[(\sigma k)_{1} - \lambda /\mu ]}
{v + i[(\sigma k)_{1} - \lambda /\mu ]}
\prod^{N}_{j=1}\frac
{u - i[(\sigma k)_{1} - (\sigma k)_{j}]}
{u + i[(\sigma k)_{1} - (\sigma k)_{j}]}.
\label{eq:p}\end{equation}
As (\ref{eq:p}) holds for any $\sigma \in S_N $,
$(\sigma k )_1$ will take $k_1, k_2,\cdots, k_N $ respectively.
So (\ref{eq:p}) represents $N$ distinct equations.

Taking the logarithm of (\ref{eq:p}), we have a system of coupled
transcendental equations,
\begin{equation}
k_{j} = I_{j}(\frac{2\pi }{ L}) - \frac{2}{L}\left[
\tan^{-1}\left(\frac{ k_{j} - \lambda /\mu }{v}\right)
+\sum^N_{l=1}\tan^{-1}\left(\frac{ k_{j} - k_{l}}{u}\right)\right]
\label{eq:q}\end{equation}
where the choice of range for $\tan^{-1}$ is $(-\pi, \pi)$,
and $I_{j}$ takes integer values or half integer values
accordingly as $N$ is even or odd. (\ref{eq:q}) is the secular
equation for the spectrum and the $\{ I_{j} \}$ play the role of the
quantum numbers. As a result of periodicity, the total momentum is quantized
i.e.  $K= n(\frac{2\pi}{L})$, $ n$ is an integer. In particular, when
$\mu =1$, the secular equation recover the known result \cite{Yang}
of an $N+1$  particles . From (\ref{eq:h}) we learned that eigenstates
related to $\{ k_{1}, k_{2},\cdots, k_{N} \}$ and those related to
$\{ \lambda, k_{2},\cdots, k_{N} \}$, $\{ k_{1}, \lambda,\cdots, k_{N} \}$
etc. have the same energy if $\mu =1 $. However, the ($N+1$) fold degeneracy
is broken by the replacement of a particle by an impurity.

Although the transcendental equation is difficult to solve, we are able to
do some meaningful calculations for $N \rightarrow \infty $,
i.e.  thermodynamic limit.
Suppose that $ \{k_{1}, k_{2},\cdots, k_{N} \}$ is a
self-consistent solution set of (\ref{eq:q}). Consider the function
\begin{equation}
I(k) = \frac{1}{2\pi}\left\{ kL +
2\left[\tan^{-1}\left(\frac{k - \lambda/\mu }{v} \right)
+\sum^N_{l=1}\tan^{-1}\left(\frac{k - k_{l}}{u}\right)\right]\right\},
\label{eq:r}\end{equation}
which is a monotonously increasing function of $k$. Clearly, when
$I(k)$ passes through one of the quantum numbers $I_{j}$,
the corresponding $k$ is equal to $k_{j}$, one of the roots.
However, there may exist some integer
(or half integer) values for which the corresponding $k$ is not in
the set $\{k_{1}, k_{2}, \cdots, k_{N} \}$,
then such a $k$ is called a `hole'.
A smooth, positive-definite density (per length) describing the
distribution of roots and holes is naturally defined as
$ \rho (k)=\displaystyle \frac{1}{L} \frac{ dI(k)}{dk} $.
Differentiating (\ref{eq:r}) we get an
expression for $\rho (k)$. In the thermodynamic limit (i.e.
sufficiently large $N$) one can
make a replacement, namely,
$\lim_{N\rightarrow \infty}\sum^N_{i=1}f(k_{i}) = \int dk\rho (k)f(k)
-\sum_{j=1}^{m}f(h_{j})$, where
$h_{1}, h_{2},\cdots, h_{m}$ are positions of holes. Then we have
\begin{equation}
\rho(k) = \frac{1}{2\pi} + \frac{1}{\pi L}
\left[\frac{v}{v^{2} + (k-\lambda/\mu )^{2}}
-\sum_{j=1}^{m}\frac{u}{ u^{2} + (k-h_{j} )^{2}}
+\int dk' \rho(k')\frac{u}{u^{2} + (k -k')^{2}} \right]
\label{eq:s}\end{equation}
This can be solved by means of the technique of Fourier transformation.
After some calculations, one gets the density of roots and holes as a
Fourier integral. Then the energy per length of the system is obtained as
\begin{equation}  
E =\frac{\lambda^{2}}{2\mu L} + \frac{1}{4\pi}\int\int
\frac{k^{2} e^{-i\xi k }}{1 - e^{-u|\xi |}}\left[ \delta (\xi )
+ \frac{1}{L}\left(e^{-v|\xi | + i\xi \lambda/\mu }
- \sum_{j=1}^{m}e^{ i\xi h_{j}}\right)\right] d\xi dk.
\label{eq:t}\end{equation}
for repulsive case $u, v >0 $.

This work is supported by NSFC and NSF of Zhejiang province.

\end{document}